\documentclass[preprint,aps,tightenlines,floatfix,showpacs,byrevtex,nofootinbib]{revtex4}
\usepackage{epsfig}

\def\be{\begin{equation}}
\def\ee{\end{equation}}
\def\mpi{m_{\pi}}

\begin{document} 

\preprint{\vbox{\noindent\null \hfill ADP-02/68-T508\\ 
                         \null \hfill hep-ph/0203023 \\
}}

\title{Simple Quark Model with Chiral Phenomenology}
\author{ I.C. Cloet\footnote{icloet@physics.adelaide.edu.au}, \ 
D.B. Leinweber\footnote{dleinweb@physics.adelaide.edu.au} \ and
A.W. Thomas\footnote{athomas@physics.adelaide.edu.au}}

\affiliation{Special Research Centre for the Subatomic Structure of 
             Matter and      \\
             Department of Physics and Mathematical Physics, University of  Adelaide,
             SA 5005, Australia} 

\begin{abstract}
We propose a new approach to the determination of hadronic observables
in which the essential features of chiral symmetry are combined with
conventional constituent quark models.  To illustrate the approach, we
consider the simple quark model in the limit of SU(3) flavour symmetry
at the strange quark mass. The comparison with data is made after an
analytic continuation which ensures the correct leading nonanalytic
behaviour of chiral perturbation theory.  The approach not only gives
an excellent fit for the octet baryon magnetic moments but the
prediction for the $\Delta^{++}$ magnetic moment is also in good
agreement with current measurements.
\end{abstract}


\maketitle
\newpage

Quark models have traditionally suffered from two major 
shortcomings \cite{Leinweber:2001ui,Leinweber:1999nf,L}.
First they have omitted the effects of the pion and kaon clouds which
give rise to non-analytic behaviour in the quark mass. 
Second, implicit in the simple constituent quark picture is the idea
that the contribution made by a quark to a hadronic observable is 
independent of its environment.
For example, the contribution of the $u$ quark to the magnetic
moment of the neutron ($(udd)$) and the $\Xi^0$ ($(uss)$) is usually
taken to be the same.
However, for the quark masses considered in lattice QCD calculations,
these environment effects are easily
observed \cite{L}. Perhaps the clearest indication of a problem is the
enormous violation of charge symmetry in the constituent quark
masses quoted by the Particle Data Group \cite{PDT}, with $M_u=338$ MeV
and $M_d=322$ MeV, differing by an unacceptable 5\%.

Recent studies of the variation of hadron properties as a function of
(current) quark mass within lattice QCD \cite{LLT,HLT,Detmold:2001jb}
have led to new insights into
hadron structure, which suggest a relatively simple approach to
overcoming these problems, while avoiding the complexities of
fully-fledged chiral quark models \cite{Theberge:1982xs,Manohar:1983md}. 
In particular, these studies have
revealed the following behaviour with quark mass \cite{DLT}:
\begin{itemize}
\item In the region of current quark masses $m > 60$ MeV or so ($m_\pi$
greater than typically 400-500 MeV)
hadron properties are smooth, slowly varying
functions of something like a constituent quark mass, 
$M \sim M_0 + c~m$ (with $c \sim 1$).
\item Indeed, $M_N \sim 3 M, M_{\rho, \omega} \sim 2 M$
and magnetic moments behave like $1/M$.
\item As $ m$ decreases below 60 MeV or so, chiral symmetry leads to
rapid, non-analytic variation.
\end{itemize}

The speed with which rapid chiral variations are suppressed above 60 MeV
or so suggests that this is the region in which constituent quark models
should be most appropriate. The connection to the physical world,
including quantitative fits to experimental data, should be undertaken
after chiral extrapolation in a manner consistent with the general
constraints of chiral perturbation theory ($\chi$PT).
  
With this in mind, we explore the utility of employing a quark model
in the region of heavier quark mass and then extrapolating
to the physical world with an analytic 
continuation of $\chi$PT. This 
extrapolation function builds in the leading nonanalytic (LNA) behaviour
of $\chi$PT and therefore explicitly incorporates the pion and kaon contributions 
in extrapolating to the chiral regime. As a first example we illustrate this 
procedure with the simple SU(6) constituent quark model (CQM) in the calculation
of octet baryon magnetic moments. We begin with the CQM in the 
limit of SU(3)--flavour symmetry where all three quarks have the same mass --
taken to be the strange quark mass. These baryon magnetic moments, determined 
at large quark masses where constituent quark degrees of freedom are manifest
\cite{DLT}, are then analytically continued to the physical mass regime. 
The resulting description of the experimental octet baryon magnetic 
moments is excellent and when applied to the charged $\Delta$ baryons the model
also produces values in good agreement with current data.

In what follows we first present the extrapolation technique used 
to link the baryon moments calculated near the limit of SU(3)--flavour
symmetry to the physical world. We then present the details of the model
(referred to as AccessQM\footnote{The name indicates the mathematical
origins of the model: Analytic Continuation of the Chiral Expansion for
the SU(6) Simple Quark Model.}) and apply it to the baryon octet.

The extrapolation technique utilized here to link baryon 
magnetic moments at large 
quark masses with the physical mass regime has only recently been exploited in 
lattice QCD. Here we extend the 
previous approach \cite{LLT,HLT} incorporating pion cloud contributions
\be
\mu(\mpi)=\frac{\mu_{0}}{1-\frac{\chi_{\pi}}{\mu_{0}}{m_{\pi}}+
\beta {m_{\pi}}^2}~~,
\label{Pion}
\ee
to include the kaon cloud. In order to place the 
leading non-analytic (LNA) kaon contribution in the denominator, 
we replace ${\chi_K}m_K$ by ${\chi_K}(m_K - m_K^{(0)})$,
where $m_K^{(0)}$ is the kaon mass in the SU(2) chiral limit 

\be
\mu(\mpi)=\frac{\mu_{0}}{1-\frac{\chi_{\pi}}{\mu_{0}}{m_{\pi}}-\frac{\chi_K}
{\mu_{0}}{(m_K-m_K^{(0)})}+\beta {m_{\pi}}^2}~~ .
\label{Pade}
\ee
We stress that $\chi_{\pi}$ and $\chi_{K}$ are model independent 
constants fixed by chiral perturbation theory (see Table~\ref{table:CHI}) and
only ${\mu}_0$ and $\beta$ are fit parameters. Further, using the Gell
Mann-Oakes-Renner relation for the pion and kaon masses, one has
\be
{m_K}^2 = {m_K^{(0)}}^2 + \frac{1}{2} {m_{\pi}}^2~~,
\ee
for fixed strange quark mass, with 
\be
m_K^{(0)} = \sqrt{(m_K^{\mathrm{phys}})^2 - 
\frac{1}{2}(m_{\pi}^{\mathrm{phys}})^2}~~.
\ee
%

%
%
%
\begin{table}[tbp]
\begin{center}
\begin{tabular}{ccccccccc}
\hline
\hline
~~~~~~~~~~&~~~~~$p$~~~~~&~~~~~$n$~~~~~&~~~~~$\Lambda$~~~~~&~~~~~$\Sigma^{+}$~~~~~&~~~~~$\Sigma^{0}~~~~~$
&~~~~~$\Sigma^{-}$~~~~~&~~~~~$\Xi^{0}$~~~~~&~~~~~$\Xi^{-}$~~~~~\\
\hline
$\chi_{\pi}$  & -4.41   &4.41      & 0              &-2.46              &0                  &2.46         &-0.191    &0.191\\
$\chi_K$      &-1.71    &-0.133    &1.47            &-3.06              &-1.47              &0.133        &3.06      &1.71\\
\hline
\hline
\end{tabular}
\end{center}
\caption{The baryon chiral coefficients, $\chi_i$, for the spin-1/2 octet.
Coefficients are from Ref. \protect\cite{JLM}, 
with one-loop corrected values for $D=0.61$ and $F=0.40$. Note here we have
suppressed the 
kaon loop contribution in the calculation of the $\chi_K$ by using
$f_{K}=1.2~f_{\pi}$.}
\label{table:CHI}
\end{table}
%
Motivated by the success of recent studies of the behaviour
of hadron properties calculated using lattice QCD 
as a function of quark mass, we now consider an amalgamation of the CQM
with the techniques of chiral extrapolation developed there.
In particular, we take as the {\em
input} for the extrapolation to the chiral limit, the CQM for the baryon
magnetic moments in the
SU(3) limit. At sufficiently large quark mass ($m_u=m_d=m_s$ near the
physical strange quark mass) chiral loop contributions should be
suppressed. Since the extrapolation function involves two
parameters, we need two input values for each baryon and these are
obtained by uniformly shifting the masses of the {\em u} and {\em d}
quarks slightly below and then slightly above the physical strange quark
mass
\begin{eqnarray}
M_{\mathrm{1}} = M_{s} - {\Delta}M~~, \nonumber \\
M_{\mathrm{2}} = M_{s} + {\Delta}M~~, \hspace*{2pt}
\label{down}
\end{eqnarray}
where we consistently use a capital $M$ for a constituent quark mass and
$m$ for a current quark mass, throughout this paper.

In Eqs.~(\ref{down}) ${\Delta}M$ and $M_s$ are input parameters. 
The magnetic moments of the baryons are simply related to the 
constituent quark masses via 
\begin{center}
\begin{tabular}{ccccccccc}
${\mu_p}$         &=&$(4{\mu_u}-{\mu_d})/3$&~~,~~~~~~~~&${\mu_n}$          &=& $(4{\mu_d }-{\mu_u})/3$~~,\\
${\mu_{\Sigma^+}}$&=&$(4{\mu_u}-{\mu_s})/3$&~~,~~~~~~~~&${\mu_{\Sigma^-}}$ &=& $(4{\mu_d }-{\mu_s})/3$~~,\\
${\mu_{\Xi^0}}$   &=&$(4{\mu_s}-{\mu_u})/3$&~~,~~~~~~~~&${\mu_{\Xi^-}}$    &=& $(4{\mu_s}-{\mu_d })/3$~~,\\
${\mu_{\Lambda}}$ &=&${\mu_s}$~~,~~~~~~~~~ &~~~~~~~~~~&&\\
\end{tabular}
\end{center}
with 
\begin{center}
$
{\mu_u} = \frac{2}{3} \frac{M_N}{M_u}~{\mu_N}~,~~~~ {\mu_d } = 
-\frac{1}{3} \frac{M_N}{M_d }~{\mu_N}~,~~~~   
{\mu_s} = -\frac{1}{3} \frac{M_N}{M_s}~{\mu_N}~,
$
\end{center}
where $M_N$ is the nucleon mass and $M_{u}=M_{d}=M_i$,
as discussed above.

To fit Eq.~(\ref{Pade}), which is a function of $m_{\pi}$, to the two 
magnetic moments obtained with $M_u=M_d=M_i~(i=1,2)$, 
we relate the pion mass to the constituent quark mass via 
the current quark mass. Chiral symmetry provides
\be
\frac{m_q}{m^{\mathrm{phys}}_q}=\frac{m_{\pi}^2}
{(m_{\pi}^{\mathrm{phys}})^2}~~,
\label{equal}
\ee
where $m^{\mathrm{phys}}_q$ is the quark mass associated 
with the physical pion mass, $m_{\pi}^{\mathrm{phys}}$.
{}From lattice studies, we know that this relation 
holds well over a remarkably large regime of pion masses, 
up to $m_{\pi} \sim 1$ GeV. The link between constituent 
and current quark masses is provided by 
\be
M = M_{\chi} + c~m_q~~,
\ee
where $M_{\chi}$ is the constituent quark mass in the 
chiral limit and $c$ is of order 1. Using Eq.~(\ref{equal})
this leads to
\be
M = M_{\chi} + 
\frac {c~m_q^{\mathrm{phys}}}{(m_{\pi}^{\mathrm{phys}})^2}~m_{\pi}^2~~.
\ee 

\noindent The link between $M_i$ of Eqs.~(\ref{down}) and $m_{\pi}$ is thus provided by
\be
m_{\pi \hspace{1.5pt} {i}}^2=(m_{\pi}^{\mathrm{phys}})^2~ \frac {M_{i}
-(M_{s}-c~m_{s}^{\mathrm{phys}})} {c~m^{\mathrm{phys}}_q} 
\hspace{15mm} (i = 1, 2)~~,
\label{opt} 
\ee
where $M_{s}-{c~m^{\mathrm{phys}}_s}=M_{\chi}$ 
encapsulates information on the constituent 
quark mass in the chiral limit, and $c~m^{\mathrm{phys}}_s$ 
provides information on the strange current quark mass.
We use the ratio 

\be
\chi_{sq}= \frac {m_s^{\mathrm{phys}}}{m_q^{\mathrm{phys}}}=24.4 \pm 1.5~~,
\ee

\noindent provided by $\chi$PT \cite{Leutwyler} to express the light current quark mass, 
$m_q^{\mathrm{phys}}$, in terms of the strange current quark mass, $m_s^{\mathrm{phys}}$,
in Eq.~(\ref{opt}).

In summary the AccessQM requires three input parameters,\\
1) $M_{s}\hspace{21pt}$ -- the strange constituent quark mass, to determine the 
SU(3)-flavour limit.\\
2) $c~m^{\mathrm{phys}}_s$ -- the strange current quark mass, if $c \sim 1$.\\
3) ${\Delta}M$\hspace{15pt} -- the spacing about the SU(3)--flavour 
limit, needed to determine the  shift of the \newline 
\hspace*{21.4mm} two magnetic moments away from this limit.

The exactness
of the SU(3) symmetry is determined through the value of 
${\Delta}M$. In the limit ${\Delta}M \to 0$,
one is effectively using the magnitude and slope predicted 
by the CQM in fitting the extrapolation function.
Our conclusions are not sensitive to the choice of this parameter over the range
(0,50] MeV and we fix ${\Delta}M$ at 20 MeV.

We now fit Eq.~(\ref{Pade}) to the
two baryon magnetic moments given by the constituent quark model, one either side of the SU(3) limit, 
allowing an extrapolation 
back to the physical mass regime. This fit is
easily accomplished as we have two equations, 
given by the extrapolation function evaluated
at each $m_{{\pi}\hspace{1.5pt}i}$ (i.e. $M_i$) and two 
unknowns ${\mu}_0$ and $\beta$ -- the two fit parameters.
We then simply solve for ${\mu}_0$ and $\beta$ simultaneously, 
where the positive root provides the smooth, non-singular extrapolation.


There are a number of approaches one could take with 
which to report the theoretical predictions 
made by the AccessQM. One would be to simply 
substitute in the excepted values for $M_{s}$ and
$c~m^{\mathrm{phys}}_{s}$ then report the predicted magnetic moments
of the octet. However as these quantities are 
only approximately known, we choose to do an optimization 
over the octet. 
We choose to minimize the RMS deviation 
between the theoretical and 
experimental magnetic moments of the octet and we denote this 
optimization function by \footnote{We omit the $\Sigma^0$ moment
from the $\chi^2$ as it has not been experimentally measured.} 
\be
\chi^2 =  {\frac{1}{7} \sum_{i=1}^7 {(\mu_i-\mu_i^{\mathrm{exp}})^2}}~~.
\ee

The values returned for $M_s$ 
and $c~m^{\mathrm{phys}}_s$ are 565 MeV and  144 MeV respectively, 
with $\chi=$ 0.051 $\mu_N$. This provides some {\em a posteriori}
justification for the approach taken within the AccessQM,
as the values obtained for $M_{s}$ and  
$m^{\mathrm{phys}}_s$ (taking the preferred vale of $c$ = 1) 
lie well within the range of their expected values. 
Table~\ref{table:LOT} provides a summary of the AccessQM 
predictions compared to experiment.
 
To do a direct comparison between the CQM and the AccessQM   
we perform an analogous optimization for the CQM. We choose 
to optimize the three constituent masses $M_s$, $M_u$ and $M_d$
subject to the charge symmetry constraint that  $M_u$ and $M_d$ should be equal
to within one percent \cite{Miller:iz}. We find 
$M_{u}$ = 345 MeV, $M_{d}$ = 342 MeV and $M_{s}$ = 538 MeV
with a RMS deviation of ${\chi} = 0.122~\mu_N$. The resulting magnetic 
moments are given in Table~\ref{table:LOT}. We see that 
the AccessQM provides more than a factor of 2 reduction in the 
RMS deviation of theory from experiment. 
This gives a good indication for the need to incorporate  
meson cloud effects into conventional constituent quark models.
\begin{table}[p]
\begin{center}
\begin{tabular}{lccc}
\hline
\hline
Baryon~~~~~~~~~~&~~~~~Quark Model~~~~~&~~~~~AccessQM~~~~~  &~~~~~Experiment~~~~~\\
\hline
$p$             &~2.724               &~2.765~   &~2.793\\
$n$             &-1.826               &-1.930~   &-1.913\\
$\Lambda$       &-0.582               &-0.591~   &~~~~-0.613(4)\\
$\Sigma^{+}$    &~2.613               &~2.532~   &~~~~~~2.458(10)\\
$\Sigma^{-}$    &-1.027               &-1.079~   &~~~~~-1.160(25)\\
$\Xi^{0}$       &-1.381               &-1.247~   &~~~~~-1.250(14)\\
$\Xi^{-}$       &-0.471               &-0.583~   &~~~~~~~-0.6507(25)\\
\hline
$M_{s}$               &  538             &565~~~~~  &${\approx}~550$\\
$c~m^{\rm{phys}}_{s}$             & $(M_{u} = 346)$  &144~~~~~  &$75~to~170$ (at 2 $\rm GeV^2$)\\
$c~m^{\rm{phys}}_{q}$ & $(M_{d} = 342)$  &5.90    &$~~2~to~7$~~~(at 2 $\rm GeV^2$)\\
\hline
${\chi}~({\mu_N})$ &0.122 &0.051&\\
\hline
\hline
\end{tabular}
\end{center}
\caption{The optimized AccessQM and CQM predictions for the magnetic moments 
of the spin--1/2 baryon octet. The values for the experimental magnetic moments
are taken from Ref. [3]. Note, the quoted value for $c~m^{\rm{phys}}_{q}$ is just
$c~m^{\rm{phys}}_{s}/{\chi_{sq}}$.}
\label{table:LOT}
\end{table} 
\begin{figure}[tbp]
\begin{center}
{\epsfig{file=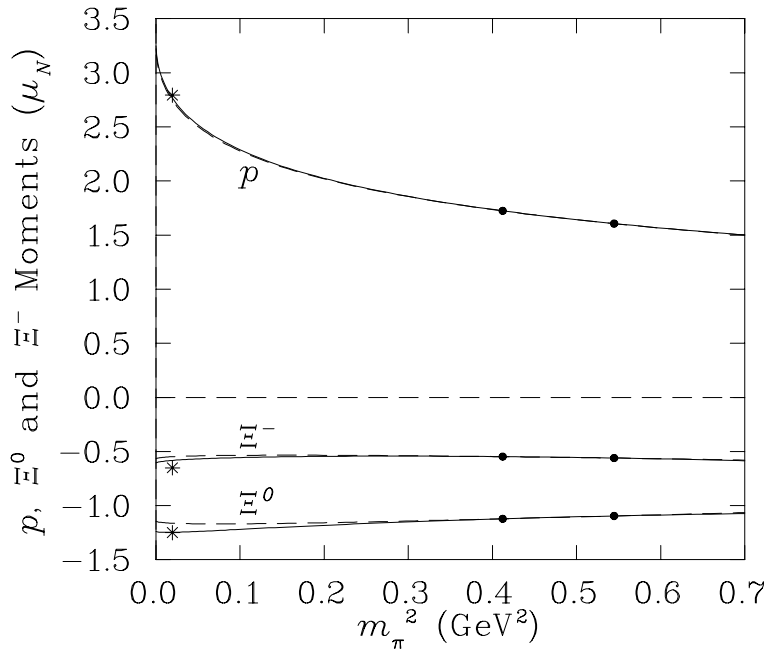, height=10cm, angle=90}}
\caption{The extrapolation function fit for the $p,\ \Xi^-$ and $\Xi^0$ 
magnetic moments. The experimental values are shown as 
asterisks ($\ast$) and the magnetic moments either side of 
the SU(3)-flavour limit as dots ($\bullet$).
The extrapolation including only the chiral behaviour associated with
the pion cloud is shown by the dashed lines.}
\label{fig:prot}
\end{center}
\end{figure}
\begin{figure}[tbp]
\begin{center}
{\epsfig{file=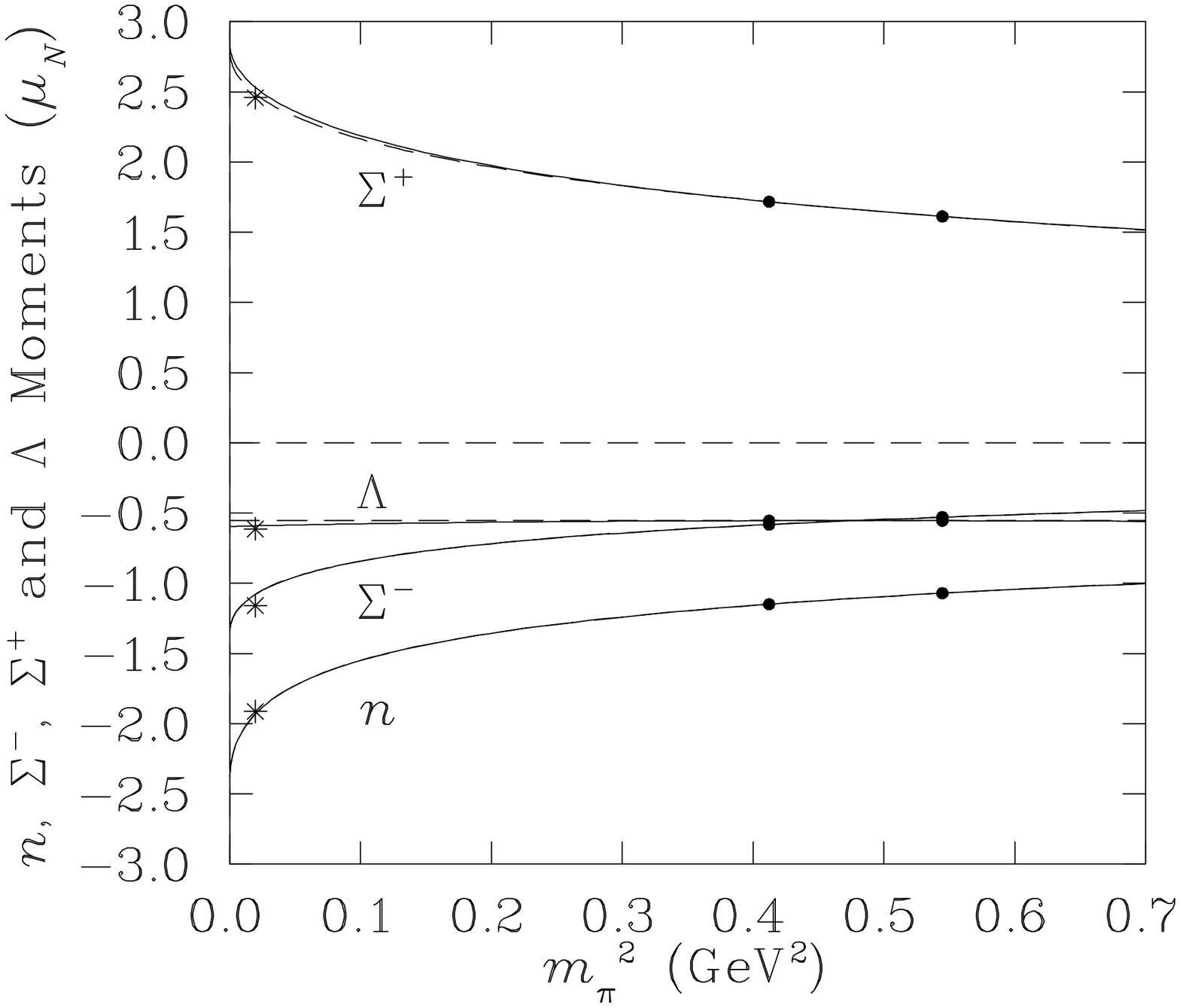, height=10cm, angle=90}}
\caption{The extrapolation function fit for the $n,\ \Sigma^-, 
\Sigma^+$ and $\Lambda$ magnetic moments. The experimental values are shown 
as asterisks ($\ast$) and the magnetic moments either side 
of the SU(3)-flavour limit as dots ($\bullet$). 
The extrapolation including only the chiral behaviour associated with
the pion cloud is shown by the dashed lines.
Note that the addition of the kaon-cloud results in 
negligible changes for the neutron and $\Sigma^-$ magnetic moments.}
\label{fig:neut}
\end{center}
\end{figure}

Figs.~1 and 2 show the behaviour of the analytic continuation of $\chi$PT fitted 
to the two magnetic moments, one either 
side of the SU(3)-flavour limit. These baryon magnetic 
moments near the SU(3)-limit
are indicated by a dot ($\bullet$) and experimental values for 
the baryon magnetic moments are given, at the 
physical pion mass, by an asterisk ($\ast$). 
To obtain the AccessQM magnetic moment prediction one simply reads off the
value of the extrapolation function at the physical pion mass. 
Note, we have also included a fit where only the chiral
behaviour associated with the pion-cloud is considered -- this is indicated  
by the dashed line. These results are obtained 
by using the same values for the input parameters, only this time setting the 
respective $\chi_K$'s to zero in Eq.~(\ref{Pade}) 
prior to determining $\mu_0$ and $\beta$. 
It is evident, from Figs. 1 and 2,
that the role of the kaon-cloud is slight 
over the octet. Indeed, it only plays a significant role for the 
$\Lambda$, $\Xi^0$ and  $\Xi^-$  magnetic moments.

A longstanding problem of the CQM is its prediction of the 
$\Xi^-$/$\Lambda$ magnetic moment ratio. The CQM 
predicts this ratio is given by 
\be
\frac{{\mu_{\Xi^-}}}{{\mu_{\Lambda}}} =
\frac{1}{3}\left(4-\frac{\mu_d}{\mu_s}\right)~~,
\ee
which becomes
\be
\frac{{\mu_{\Xi^-}}}{{\mu_{\Lambda}}} = 
\frac{1}{3}\left(4-\frac{M_s}{M_d}\right)~~,
\ee
and is therefore less than 1, as $M_{d} < M_{s}$. Experiment 
places this ratio at 1.06(1). The addition of 
the meson cloud gives us the opportunity of solving this problem, 
and indeed we find from Table~\ref{table:LOT} a ratio of 0.99 for the
AccessQM, compared with
0.81 for the CQM. While this is a significant improvement,
there is a residual disagreement, suggesting  
the need to replace the very simplest CQM with 
something a little more sophisticated \cite{Morpurgo:2001mc}.       

We have shown that using the very simplest CQM, with all three quark
masses near the physical strange quark mass, and extrapolating to
the physical mass regime using an  
analytic continuation of $\chi$PT which ensures the correct LNA behaviour in the
chiral limit, does offer a considerable improvement to the theoretical
predictions of the spin--1/2 baryon octet magnetic moments, 
as compared to those of the CQM alone.
The results indicate the importance of 
incorporating the meson cloud contribution in any calculation 
of baryon magnetic moments as well as the need to 
accommodate the hadronic environment of the constituent quarks.

This work serves to introduce the idea that one should merge
the general class of constituent quark models with known chiral 
properties of hadronic observables. While the results 
presented here display great promise, there is a demonstrated need for 
further refinement. For example,
one could explore the possibility that 
decuplet baryon intermediate states make an important contribution 
in the chiral extrapolation. For the decuplet itself there is limited
experimental data, but for the $\Delta^{++}$, which is known to lie in
the range [3.7,7.5] $\mu_N$ \cite{PDT}, 
with the latest experimental measurement
yielding $4.52 \pm 0.50 \pm 0.45$ \cite{Bosshard:zp},
the application of the 
model outlined here yields a value of 4.67 $\mu_N$ -- in good
agreement with the data. The predictions for the remaining 
charged $\Delta$ baryons are $\mu_{\Delta^+} = 2.29~\mu_N$ and 
$\mu_{\Delta^-} = -2.50~\mu_N$.
It will also be possible to tune these models
to data at larger quark masses from the new generation of lattice QCD
simulations now underway. Finally, there is also an important opportunity  
to refine the constituent quark model itself, where
spin-dependent interactions \cite{DGG,PB} between quarks can give rise to 
important contributions \cite{CGI,IK}. 
It is our hope that the ideas presented here
will lead to a new appreciation of the role of the
constituent quark model in modern hadron phenomenology in which there
is no longer a conflict with the constraints of chiral perturbation
theory.

\section*{Acknowledgement}
We thank Ross Young for helpful discussions.
This work was supported by the Australian Research Council 
and the University of Adelaide.

\end{document}